\documentclass[12pt]{article}
\usepackage{amsmath}
\usepackage{enumerate}
\usepackage{natbib}
\usepackage{url} 

\date{July 8, 2016}

\newcommand{\blind}{0}

\usepackage{natbib}
\usepackage{lscape,rotating}
\usepackage{float,epsfig}
\usepackage{subfigure}
\usepackage{setspace}

\usepackage{amsmath} 
\usepackage{amsfonts}
\usepackage{amssymb}

\newtheorem{lemma}{Lemma}

\usepackage{lscape,rotating}
\usepackage{float,epsfig}
\usepackage{subfigure}
\usepackage{setspace}
\usepackage{amsmath}
 \usepackage{amsmath,bm} 
\usepackage{amsfonts}
\usepackage{amssymb}

\newenvironment{proof}[1][Proof]{\begin{trivlist}
\item[\hskip \labelsep {\bfseries #1}]}{\end{trivlist}}
\newcommand{\ev}{\bm{e}}

\newcommand{\xv}{\bm{x}}

\newcommand{\yv}{\bm{y}}
\newcommand{\wv}{\bm{w}}
\newcommand{\zv}{\bm{z}}

\newcommand{\deltav}{\bm{\delta}}
\newcommand{\Deltav}{\bm{\Delta}}

\newcommand{\muv}{\bm{\mu}}
\newcommand{\phiv}{\bm{\phi}}

\newcommand{\etav}{\bm{\eta}}
\newcommand{\piv}{\bm{\pi}}

\newcommand{\thetav}{\bm{\theta}}
\newcommand{\alphav}{\bm{\alpha}}
\newcommand{\betav}{\bm{\beta}}

\newcommand{\zerov}{\bm{0}}
\newcommand{\onev}{\bm{1}}

\begin{document}

\def\spacingset#1{\renewcommand{\baselinestretch}%
{#1}\small\normalsize} \spacingset{1}

%%%%%%%%%%%%%%%%%%%%%%%%%%%%%%%%%%%%%%%%%%%%%%%%%%%%%%%%%%%%%%%%%%%%%%%%%%%%%%

\if0\blind
{
  \title{\bf An efficient  multiple imputation algorithm for control-based and delta-adjusted pattern mixture models using SAS}
  \author{Yongqiang Tang \hspace{.2cm}\\
   Shire, 300 Shire Way, Lexington, MA 02421, USA\footnote{ to appear in Statistics in Biopharmaceutical research}}
  \maketitle
} \fi

\if1\blind
{
  \bigskip
  \bigskip
  \bigskip
  \begin{center}
    {\LARGE\bf An efficient  multiple imputation algorithm for control-based and delta-adjusted pattern mixture models}
\end{center}
  \medskip
} \fi

%\author[shire]{Yongqiang Tang,}\ead{yongqiang\_tang@yahoo.com}
%\address[shire]{Shire, 300 Shire Way, Lexington, MA 02421, USA}

\bigskip
\begin{abstract}
In clinical trials, mixed effects models for repeated measures (MMRM) and pattern mixture models (PMM) 
are often used to analyze longitudinal continuous outcomes.
We describe a simple missing data imputation algorithm for the MMRM that can be easily implemented in standard statistical software packages
such as SAS PROC MI. 
We explore the relationship of the missing data distribution in the control-based and delta-adjusted PMMs
with that in the MMRM, and suggest an efficient imputation algorithm for these PMMs.
 The unobserved values in  PMMs can be imputed  by subtracting the mean difference in the posterior predictive
distributions of missing data from the imputed values in MMRM. 
We also suggest a modification of the copy reference imputation procedure to avoid the possibility that after dropout, subjects from the active treatment arm will have  
better mean response trajectory than subjects  who stay on the active treatment.
The proposed methods are illustrated by the analysis of  an  antidepressant trial.

\end{abstract}

\noindent%
\vfill

\newpage
\spacingset{1.45} % DON'T change the spacing!
\section{Introduction}\label{sec:intro}

Missing data are unavoidable  in clinical trials, and can potentially result in biased treatment effect estimates. 
The primary analysis generally assumes a missing at random (MAR) mechanism. 
Suppose two subjects are identical (same treatment, same historical outcome) until a certain time point, and
subject $A$ discontinues from the study after that time point, but  subject $B$ remains on the treatment. The MAR  mechanism
implies that the future statistical behaviors of the two subjects are expected to be the same. 
% no matter when subject $A$ discontinues the treatment. 
This assumption may be unrealistic in some applications. For example, 
in a trial investigating a symptomatic treatment, the treatment benefit may disappear rapidly after discontinuation of the treatment, 
and dropouts and completers are unlikely to have the same statistical behaviors. 

Recent regulatory guidelines \citep{ICHE9,chmp:2010,NRC:2010} emphasize the importance of sensitivity analyses to assess the robustness 
of the trial result under the missing not at random (MNAR) assumption.
The  pattern-mixture models (PMM) have become increasingly popular in sensitivity analyses for handling longitudinal clinical
data with nonignorable missingness because the underlying missing data assumptions are easy to understand and interpret 
\citep{lu:2014a}. Two types of PMMs  commonly used as MNAR sensitivity analyses in confirmatory trials are 
the control-based   and  delta-adjusted PMMs  \citep{little:1996, 2013:carpenter,  ratitch:2013}. 
These PMMs provide {\it de facto} (effectiveness) estimands of the treatment effect, which measure the effect of the drug as
actually taken using all randomized subjects under the intent to treat (ITT) principle \citep{ 2013:carpenter, ayele:2014}.
Recently  an addendum to ICH E9 was proposed relating to estimands and sensitivity analyses, and 
an expert group meeting sponsored by Statisticians in the Pharmaceutical Industry (PSI)  was devoted to  this topic 
 prior to the release of the addendum \citep{phillips:2016}. The estimands from PMMs  and MMRM  correspond respectively to estimands  $2$ and $3$ illustrated at the meeting
 \citep{phillips:2016}.
These PMMs assume that  the benefit among subjects in the active arm disappears or diminishes after treatment discontinuation, 
and therefore generally yield more conservative  treatment effect estimates  than the MMRM.

PMMs are often  implemented via multiple imputation (MI). The simplest imputation algorithm is the sequential regression imputation (SRI)
for monotone data, i.e.  no missing data prior to dropout  \citep{little:1996, ratitch:2013, lu:2014a}.  \cite{tang:2015, 2016:tang} proposed  Markov chain Monte Carlo (MCMC) algorithms
via monotone data augmentation (MDA)  for missing data imputations in  mixed effects models for repeated measures (MMRM) and 
a class of PMMs that assume the same observed data distribution as the MMRM.  The MDA algorithm is a  collapsed Gibbs sampler \citep{liu2:1994}, in which 
the missing data $\yv_{i\text{m}_2}$'s after dropout are integrated out from the posterior distribution, and only the intermittent missing data $\yv_{\text{m}_1}$ are imputed
in the imputation I-step.
Compared to a full data augmentation (FDA) algorithm that imputes both $\yv_{i\text{m}_1}$'s and  $\yv_{i\text{m}_2}$'s in the  I-step, the MDA algorithm generally 
converges to the same stationary distribution faster with smaller autocorrelation between posterior
samples   \citep{ schafer:1997, 2016:tang}. 
%Furthermore, when the missing data distribution is complex in PMMs, sampling the model parameters from their posterior distribution  can be complicated
%in the FDA algorithm, but it is straightforward to implement the MDA algorithm  \citep{2016:tang}. 

The purpose of the paper is to propose a method that enables 
 readers to write simple and efficient computer code for missing data imputations in the MMRM, control-based PMMs and delta-adjusted PMMs by 
building on existing software packages (e.g. PROC MI in SAS, Norm package in R) that provide missing data imputation
for incomplete multivariate normal data. The proposed method will  produce the same posterior predictive distribution for the  missing data 
as   \cite{tang:2015, 2016:tang}  MDA algorithm.

Our algorithm involves four steps, and details will be explained  later in this section.
\begin{itemize}
\item[1.] Construct a prior  $p(\phiv_x,\phiv_y)=p(\phiv_x)p(\phiv_y)$   that can easily be specified in the software,
where $\phiv_x$ and $\phiv_y$ are parameters associated respectively with the distributions of the covariates $\tilde\xv_i$ and outcomes $\yv_i$.
\item[2.]  Impute missing data for MMRM (it assumes MAR) under the prior $p(\phiv_x,\phiv_y)$ using the software by pretending $\wv_i=(\tilde\xv_i',\yv_i')'$ is normally distributed.
\item[3.] Calculate the difference in the posterior mean of the missing data in MMRM and PMMs.
 It  is a function  of the posterior sample of  $\phiv_y$. %The missing data distribution is  summarized in Table \ref{deltat}, Section $4$.
\item[4.] Subtracting the difference from the MAR-based imputation yields the imputed values for PMMs. 
\end{itemize}

Section $2$ describes the imputation algorithm for MMRM (i.e. steps $1$ and $2$). Let $f(\yv_{i}|\tilde\xv_i,\phiv_y)$ and $f(\yv_{io}|\tilde\xv_i,\phiv_y)$
 denote respectively the conditional distributions of the complete outcome $\yv_i$ and observed outcome $\yv_{io}$ given the covariates $\tilde\xv_i$ in the MMRM, and $p(\phiv_y)$ the prior.
The missing values in the MMRM are imputed by pretending $\wv_i=(\tilde\xv_i',\yv_i')'$ follows a joint multivariate normal distribution with density $f(\wv_i|\phiv_x,\phiv_y)= f(\yv_i|\tilde\xv_i,\phiv_y)f(\tilde\xv_i|\phiv_x)$.
 But it does not require $\tilde\xv_i$ to be normally distributed or continuous (that is, $f(\tilde\xv_i|\phiv_x)$ is not the true density for $\tilde\xv_i$).
Since $\phiv_x$ and $\phiv_y$ have independent prior distributions, the marginal distribution of the posterior samples 
 $\phiv_y$ in this new algorithm will converge to its target distribution $f(\phiv_y|(\tilde\xv_i,\yv_{io})'s) \propto p(\phiv_y)\prod_i f(\yv_{io}|\tilde\xv_i,\phiv_y)$.

Section $3$ describes the imputation algorithm for   control-based  and delta-adjusted PMMs (i.e. steps $3$ and $4$).
 These PMMs assume the same observed data distribution as the MMRM. Therefore
the posterior distributions of $\phiv_y$  in these PMMs are the same as that in MMRM \citep{2016:tang}.
Furthermore, the missing data after dropout are assumed to be normally  distributed with the same covariance matrix as that in MMRM. 
Thus the  missing values in  PMMs can  be imputed by subtracting the mean difference from the imputed values in the MMRM. 

In the copy reference (CR) PMM, it is possible that the mean response after dropout among drug-treated subjects could be better than that among subjects who remain
on the active treatment. We propose a modification of the CR procedure in Section $3$. 

Section $4$ illustrates the proposed methods by the analysis of an antidepressant trial.
Section $5$ will  compare the proposed imputation algorithm with some existing methods.

\section{MCMC algorithms for  MMRM}
\subsection{MMRM and related MCMC algorithms}
We consider a two-arm trial, where $n$ subjects are randomly assigned to the active ($g_i=1$) or control ($g_i=0$) treatment.  
Let $\yv_i =(y_{i1},\ldots,y_{ip})'$ denote
the  outcomes at the $p$ post-baseline visits, and  $(x_{i1},\ldots,x_{id})'$ the baseline  covariates for subject $i$. We assume that $x_{ik}$'s are fully observed.
In general,  $\yv_i$'s will  be only partially observed. 
Let $r_i$ index the dropout pattern for subject $i$ according to the last observation.  A subject is  in pattern $r_i=s$  if $s$ is last visit that 
the subject has a measurement observed, and $r_i=0$ if a subject has no post-baseline assessment.
Without loss of generality, we sort the data so that subjects in pattern $s$ are arranged before subjects in pattern $t$ if $s>t$.
 Suppose after filling in the intermittent missing data, $y_{ik}$ is not missing in  the first $n_k$ subjects.

Let $\yv_{io}$, $\yv_{i\text{m}_1}$ and $\yv_{i\text{m}_2}$ denote respectively the observed data, intermittent missing data, and
missing data after dropout for subject $i$. Let $\vec{\yv}_{it}=(y_{i1},\ldots,y_{it})$. 
Then $\yv_{io}$ ($\yv_{im_1}$) is the observed (missing) 
part of $\vec{\yv}_{ir_i}$.  Let  $x_{i0}\equiv 1$, $\tilde\xv_i =(x_{i1},\ldots,x_{id}, g_i)'$ and $\xv_i =(x_{i0},\ldots,x_{id}, g_i)'$. 
Let $Y_o$, $Y_{\text{m}_1}$, $Y_{\text{m}_2}$, and $X$ denote respectively the observed outcomes, 
intermittent missing data, missing data after dropout, and covariates from all subjects.

The following MMRM is often  used as the primary analysis for  longitudinal  outcomes collected at a fixed number of time points  \citep{aiddiqui:2009}.
It assumes MAR.
\begin{equation}\label{mixed3}
\yv_i=(y_{i1},\ldots,y_{ip})' \sim N[(\alphav_{1}' \xv_i,  \ldots, \alphav_{p}' \xv_i)', \Sigma]
\end{equation} 
where $q=d+2$, and $\alphav_j=(\alpha_{j0}, \ldots,\alpha_{jd},\delta_j)'$ be a $q \times 1$  vector of the covariate and treatment effects at visit $j$.
The model includes an  unstructured treatment $\times$ visit  interaction effect, and thus allows the direct estimate of the treatment effect at each scheduled visit \citep{aiddiqui:2009}. 
  The within subject errors are modeled by an unstructured covariance.

Let $\Sigma = U^{-1} \,\Lambda  (U')^{-1}$ be the LDL decomposition of $\Sigma$, where 
 $\Lambda =\text{diag}(\gamma_1^{-1},\ldots,\gamma_p^{-1})$, and $U=$ {$\begin{bmatrix} 1 & 0 &  \ldots & 0\\
                       -\beta_{21} & 1 & \ldots &0 \\
                       & \ldots & \ldots & 0 \\
                      -\beta_{p1} & \ldots & -\beta_{p,p-1} & 1 \\
                    \end{bmatrix}$}. Then \eqref{mixed3} can be written as 
$U \yv_i \sim N[(\underline\alphav_1'\xv_i,\ldots,\underline\alphav_p'\xv_i)', \Lambda]$, or equivalently as the product of the following  regression models
\begin{equation}\label{factor} 
y_{ij}|\vec{\yv}_{i,j-1},\xv_i \sim   N\left(\thetav_j' \zv_{i,j-1}, \gamma_j^{-1}\right) \text{ for } j\leq p,
\end{equation}
 where 
 $\underline{\alphav}_{j} = (\underline \alpha_{j0}, \ldots,\underline\alpha_{jd},\underline\delta_j)' =\alphav_{j} - \sum_{t=1}^{j-1} \beta_{jt} \alphav_{t} $, 
    $\betav_j=(\beta_{j1},\ldots,\beta_{jj-1})'$, $\thetav_j=(\underline{\alphav}_{j}', \betav_j')'$,
    and $\zv_{ij} =(\xv_i',\vec{\yv}_{ij}')'$. Let $\alphav=(\alphav_1,\ldots,\alphav_p)'$.

\cite{tang:2015} considered a conjugate normal-inverse-Wishart (NIW) prior for $(\Sigma,\alphav)$. That is,  $\Sigma \sim \mathcal{W}^{-1}(A,\nu_0)$,
$\text{vec}(\alphav) |\Sigma \sim N(\text{vec}(\alphav_0),  M^{+} \otimes\Sigma)$, and the prior density  is 
 \begin{eqnarray}\label{priorsig}
 \begin{aligned}
p(\alphav,\Sigma)  \propto & p(\Sigma)  p(\alphav|\Sigma)\propto \left\{|\Sigma|^{-\frac{\nu_0+p+1}{2}}\exp\left[-\frac{1}{2}\text{tr}(A\Sigma^{-1})\right]\right\}\\
& \left\{  |\Sigma|^{-\frac{r}{2}} \exp\left[-\frac{1}{2} (\text{vec}(\alphav-\alphav_0))' ( M\otimes\Sigma^{-1}) (\text{vec}(\alphav-\alphav_0))  \right]\right\},
\end{aligned}
 \end{eqnarray}
where $\mathcal{W}^{-1}(A,\nu_0)$ denotes an inverse-Wishart distribution
 with $\nu_0$ degrees of freedom, and $p\times p$ scale matrix $A$, 
$\alphav_0$ is a $p\times q$ fixed matrix containing the prior mean of $\alphav$, 
 and $M^{+}$ is Moore-Penrose inverse of a $q\times q$ fixed matrix $M =(m_{ij})$ with rank $r$.
The prior for $q-r$  covariates is flat. 
If a covariate has a flat prior, the corresponding mean components in $\alphav_0$ and  (co)variance elements in $M$ will be set to $0$. 
For example, suppose historical information is available  only on the placebo response, and there is no baseline covariate ($d=0$).  We  may 
put a (weakly) informative prior on the intercept, and a flat prior on the treatment effect. 
Then 
$M=\begin{bmatrix} m_{11} & 0 \\
                                        0 & 0\\
                                        \end{bmatrix}$ and %$\alphav_0=\begin{bmatrix}\alpha_{01} & \ldots & \alpha_{0p}\\ 
                                                                      %0 & \ldots &0 \\\end{bmatrix}$.  
 $\alphav_0=(\alphav_{01}, \zerov_{p})$, where $\zerov_k$ is a $k\times1$ vector of zeros.
If the prior is flat on all covariates ($r = 0$), then $M = \zerov$. 
We need to pre-specify $A$, $\nu_0$, $\alphav_0$ and $M$ in the prior. For Jeffrey's prior $p(\alphav,\Sigma) \propto |\Sigma|^{-(p+1)/2}$, we have
 $\nu_0 = 0$, $A = \zerov$, $\alphav_0=\zerov$, and $M=\zerov$. We do not require $A$ to be positive definite.

\cite{tang:2015, 2016:tang} developed the MDA  (called  MDA-T) and FDA  (called  FDA-T) algorithms  for MMRM.
% which is slightly improved to reduce the amount of matrix inversion, and compared to a FDA algorithm by \cite{2016:tang}. 
They rely on the fact that $(\thetav_j,\gamma_j)$'s are independent 
in  the posterior distribution given the  augmented monotone  $(Y_o,Y_{\text{m}1})$ or complete  $(Y_o,Y_{\text{m}1}, Y_{\text{m}2})$ data since 
the corresponding likelihood  can be factored  as the product of $p$ independent likelihoods for $(\thetav_j,\gamma_j)$'s based on
\eqref{factor}, and the prior for $(\alphav,\Sigma)$ can be factored into  independent densities for $(\thetav_j,\gamma_j)$'s. 
The MDA algorithm is generally recommended in real applications because it 
converges to the same stationary distribution faster  than the FDA algorithm \citep{schafer:1997, 2016:tang}.
 Appendix \ref{mcmctang} provides a brief review of the two algorithms.

\subsection{MCMC algorithm for MMRM using SAS PROC MI}\label{mmrms}
This section describes a simple method to implement Bayesian MMRM analysis  in a software package that can generate multiple imputations for
 incomplete multivariate normal data.  We illustrate the method using SAS PROC MI since SAS is more commonly used to analyze clinical trials
in the  pharmaceutical industry.
The underlying idea is to  run a Bayesian analysis
for  $\wv_i =(\tilde{\xv}_i',\yv_i')'$ 
by pretending $\wv_i$ follows a multivariate normal distribution $\wv_i \stackrel{i.i.d}{\sim} N(\alphav_w, \Sigma_w)$, where
 $\alphav_w=\begin{bmatrix} \alphav_x  \\ \alphav_{y} \\ \end{bmatrix}$ and
  $\Sigma_w=\begin{bmatrix} \Sigma_{xx} & \Sigma_{yx}' \\ \Sigma_{yx} & \Sigma_{yy} \\ \end{bmatrix}$.
The likelihood for $(\alphav_w, \Sigma_w)$ can be decomposed as independent likelihoods for $(\alphav_x,\Sigma_{xx})$ and $(\alphav,\Sigma)$, 
$$\prod_{i=1}^n f(\wv_i;\alphav_w, \Sigma_w)=\{ \prod_{i=1}^n  f(\tilde{\xv}_i;\alphav_x,\Sigma_{xx})\}\{ \prod_{i=1}^n f(\yv_i|\xv_i;\alphav,\Sigma)\}.$$
The method yields the same posterior distribution of $(\alphav,\Sigma)$ as any valid MCMC algorithm for \eqref{mixed3} if the prior for
 $(\alphav_w, \Sigma_w)$ can be factored into independent densities as $p(\alphav_w, \Sigma_w)=p(\alphav_x,\Sigma_{xx}) p(\alphav,\Sigma)$, where
$p(\alphav,\Sigma)$ is defined in \eqref{priorsig}.  
We assume $\tilde{\xv}_i$'s are fully observed, but   $\tilde{\xv}_i$ may not  be normally distributed. 

For missing data imputation, it is convenient to reparameterize $(\alphav_w, \Sigma_w)$ based on  the LDL decomposition of  
$\Sigma_w=U_w^{-1} \,\Lambda_w  (U_w')^{-1}$, where $U_w$ is a lower triangular matrix with  all $1$'s on the diagonal, and
 $\Lambda_w =\text{diag}(\gamma_{1}^{*^{-1}},\ldots,\gamma_{q-1+p}^{*^{-1}})$. Let $-\betav_j^*$ denote the first $j-1$ elements of row $j$ in $U_w$, 
 $\underline\alpha_j^*$ the $j$-th element of $\underline{\alphav}_w=U_w \alphav_w$,
and $\thetav_j^*=(\underline\alpha_j^*, {\betav_j^*})'$. Note that $(\thetav_{j+q-1}^*,\gamma_{j+q-1}^*)$ has the same interpretation as $(\thetav_j,\gamma_j)$  defined in \eqref{factor}. 
Then
$(\alphav_x,\Sigma_{xx})$ can be expressed  a function of $\phiv_x=\{(\thetav_j^*,\gamma_j^*): j=1,\ldots,q-1\}$, and
$(\alphav,\Sigma)$ is a function of $\phiv_y=\{(\thetav_j^*,\gamma_j^*): j=q,\ldots,q+p-1\} = \{(\thetav_j,\gamma_j): j=1,\ldots,p\}$.

 Lemma \ref{sasalg} below provides the theoretical basis for the proposed algorithm, in which a NIW prior is constructed for $(\alphav_w,\Sigma_w)$. 
Its  proof  will be deferred to Appendix \ref{proof}. One needs to create the following quantities from  $A$, $\nu_0$, $\alphav_0$ and $M$ to define the NIW prior.
\begin{itemize}
\item $m_{11}$ is $(1,1)$ entry of $M=$ { $\begin{bmatrix} m_{11} & M_{12} \\ M_{21} & M_{22}\\ \end{bmatrix}$}. Note that $m_{11}^+=m_{11}^{-1}$ if $m_{11}>0$, and 
$m_{11}^+=0$ if $m_{11}=0$.
\item $A_w =$ { $\begin{bmatrix} M^* & M^*{\alphav_0^*}' \\ \alphav_0^*M^* & A+ \alphav_0^*M^*{\alphav_0^*}'\\ \end{bmatrix}$},
where  $\alphav_0^*$ is a $p\times (q-1)$ matrix containing the last $q-1$ columns of $\alphav_0$,
$M^* = M_{22}$ if $m_{11}=0$, and $M^* = M_{22}-M_{21}M_{12}/m_{11}$ if $m_{11}> 0$.
\item   $\nu_w = \nu_0+r-(q-1)$ if $m_{11}=0$, and  $\nu_w = \nu_0+r-q$ if $m_{11}\neq 0$.
\item  $\alphav_{w0} =\zerov_{q-1+p}$  if $m_{11}=0$, and $\alphav_{w0} =m_{11}^{-1} \begin{bmatrix} M_{21} \\   M_1^* \\ \end{bmatrix}$
  if $m_{11}\neq 0$, where $M_1^*= \alphav_0\begin{bmatrix} m_{11} \\ M_{21} \\ \end{bmatrix}$.
\end{itemize}
In large confirmatory trials,  a flat prior is generally put on $\alphav$ ($r=0$, $M=\zerov$), then we have
 $A_w=\begin{bmatrix} \zerov & \zerov \\ \zerov & A\\ \end{bmatrix}$,  $\nu_w = \nu_0-q+1$, $m_{11}=0$ and 
$\alphav_{w0} =\zerov_{q-1+p}$.

\begin{lemma}\label{sasalg}
Suppose in a Bayesian analysis, $(\alphav_w, \Sigma_w)$ or $(\underline\alphav_w, \Lambda_w)$ are sampled under the NIW prior  $\Sigma_w \sim \mathcal{W}^{-1}(A_w,\nu_w)$ and $\alphav_w |\Sigma_w \sim N(\alphav_{w0}, m_{11}^{+} \Sigma_w)$ 
(when $m_{11}=0$,  $f(\alphav_w|\Sigma_w)\propto \text{constant}$) by  pretending  
$\wv_i \stackrel{i.i.d}{\sim} N( \alphav_w, \Sigma_w)$.
Then \\
(a) $(\thetav_j^*,\gamma_j^*)$'s are independent in the prior, and  $(\thetav_{j+q-1}^*,\gamma_{j+q-1}^*)$ has the same prior as $(\thetav_j,\gamma_j)$ defined in \eqref{priortheta} in Appendix \ref{mcmctang}.\\
(b) $(\alphav_x,\Sigma_{xx})$ and $(\alphav,\Sigma)$ are independent in the prior.\\
(c) If  \eqref{mixed3} holds, but $\tilde\xv_{i}$ may not be normally distributed, the posterior distributions of $\phiv_y$ and $(\alphav,\Sigma)$ are the same as
 as that from the MCMC algorithms  discussed in Section $2.1$.
\end{lemma}

In SAS, MCMC sampling for incomplete multivariate normal data can be conveniently performed using PROC MI. It is flexible to specify  priors in SAS.  
 The prior on $\alphav_w$ could be flat (i.e. $m_{11}=0$, $f(\alphav_w|\Sigma_w)\propto \text{constant}$), 
and $A_w$ may not be of full rank. SAS PROC MI produces posterior samples of $(\alphav_w, \Sigma_w)$.
%Let the lower triangular matrix $G_w=\begin{bmatrix} G_{xx} & \zerov\\ G_{yx} & G_{yy} \\ \end{bmatrix}$ be Cholesky factor of $\Sigma_w=G_wG_w'$, 
%where  $G_{xx}$ is a $(q-1)\times (q-1)$ square matrix. 
One can recreate $\phiv_y= \{(\thetav_j,\gamma_j): j=1,\ldots,p\}$  from $(\alphav_w,\Sigma_w)$ based on LDL decomposition of $\Sigma_w$, and 
recreate $(\alphav,\Sigma)$  from $(\thetav_j,\gamma_j)$'s based on $\Sigma=U^{-1} \,\Lambda  (U')^{-1}$, and $\alphav= U^{-1} [\underline\alphav_1,\ldots,\underline\alphav_p]'$.
% One can also recreate $(\alphav,\Sigma)$ directly based on 
% $\Sigma = \Sigma_{yy}-\Sigma_{yx}\Sigma_{xx}^{-1}\Sigma_{yx}'= G_{yy} G_{yy}'$,
  % $\alphav^*=\begin{bmatrix} \alpha_{11} &\ldots &\alpha_{d1} &\delta_1\\
   %                                             & \ldots & \\
   %                                            \alpha_{1p} &\ldots &\alpha_{dp} &\delta_p
  %                                             \end{bmatrix}$
 % $ =\Sigma_{yx}\Sigma_{xx}^{-1}=G_{yx}G_{xx}^{-1}$, and
 % $(\alpha_{10},\ldots, \alpha_{p0})'=\alphav_y -\alphav^*\alphav_x$. 

SAS PROC MI implements  \cite{schafer:1997}  MDA algorithm (called MDA-SAS)  under Jeffrey's  prior, and a FDA algorithm (called FDA-SAS) under a general prior.
To provide further insight into the proposed method, 
we will explain in Appendix \ref{mcmcsas} why the two algorithms in SAS yield  the same posterior distribution of $\phiv_y$ as the MDA-T and FDA-T algorithms  without the use of Lemma  \ref{sasalg}.

\section{Missing data imputation in PMMs}\label{meandiffpmm}
\setlength{\parindent}{3ex}
PMMs are generally implemented via  MI. 
In MI, $m$ ($m>1$) complete datasets are imputed, and  analyzed using a standard method (e.g.  MMRM, analysis of covariance model (ANCOVA)).
The results from the $m$ complete datasets are then combined using  \cite{rubin:1987} rule.
One major challenge in the MI inference is the imputation of missing data. 
We will introduce a simple and efficient strategy for  missing data imputation in a class of PMMs that assume the same observed data distribution as  MMRM.

%The PMM stratifies subjects by pattern $r_i$, and  models the difference in distribution of $\yv_i$ over these patterns \citep{little:1995}.
 The joint distribution of $(\yv_i,r_i)$ in PMMs can be factored as
\begin{eqnarray*}
\begin{aligned}
f(\yv_i,r_i|\phiv, \piv, \xv_i)&= f(\yv_i|r_i,\phiv_y, \xv_i) f(r_i|\piv, \xv_i)\\
   &= f(\vec\yv_{ir_i}|\phiv_y, \xv_i)Q(\yv_{i\text{m}_2}|\vec\yv_{ir_i},\phiv_y, \xv_i)f(r_i|\piv, \xv_i) ,
\end{aligned}
\end{eqnarray*}
 where $f(r_i|\piv, \xv_i)$ models the marginal proportions of subjects in each pattern,  
and 
$Q(\yv_{i\text{m}_2}|\vec{\yv}_{ir_i},\xv_i,\phiv_y)$ is  the conditional distribution of $\yv_{i\text{m}_2}$ given
 $\vec{\yv}_{ir_i}=(\yv_{io}',\yv_{i\text{m}_1}')'$ and $\xv_i$. 
A common feature of these PMMs is that  the outcome $\vec{\yv}_{ir_i}$ 
before dropout has the same marginal distribution as that defined in  \eqref{mixed3}.
% i.e. $f(\vec{\yv}_{ir_i}|\phiv_y,\xv_i)\propto \prod_{j=1}^{r_i}\gamma_j^{1/2} \exp\left[-\gamma_j(y_{ij}-\nu_{ij})^2/2\right]$.
  That is, the observed data ($\yv_{io}$'s) distributions are identical in PMMs and MMRM, 
and the intermittent missing data ($\yv_{i\text{m}_1}$'s) are MAR.

In PMMs, the complete data likelihood can be written as 
 $$\left\{ \prod_{i=1}^n  f(\vec\yv_{ir_i}|\xv_i,\phiv_y)  \prod_{i=1}^n Q(\yv_{i\text{m}_2}|\vec\yv_{ir_i},\xv_i,\phiv_y)\right\}\left \{ \prod_{i=1}^n f(r_i|\xv_i,\piv)\right\}.$$
If the  prior   $\phiv_y \sim p(\phiv_y)$ is the same as specified in section $2$, and an independent 
prior is put on $\piv$, then
$\piv$ and $(\phiv_y,Y_{\text{m}_1},Y_{\text{m}_2})$ are independent in the posterior distribution, 
and 
the joint posterior distribution of $(\phiv_y,Y_{\text{m}_1},Y_{\text{m}_2})$ is given by
\begin{equation}\label{pospmm}
 \text{pos}_2(\phiv_y,Y_{\text{m}_1},Y_{\text{m}_2}|Y_o,X) \propto \text{pos}_1(\phiv_y,Y_{\text{m}_1}|Y_o,X) \prod_{i=1}^n Q(\yv_{i\text{m}_2}|\vec\yv_{ir_i},\xv_i,\phiv_y),
\end{equation}
where $\text{pos}_1(\phiv_y,Y_{\text{m}_1}|Y_o,X)\propto p(\phiv_y) \prod_{i=1}^n  f(\vec\yv_{ir_i}|\xv_i,\phiv_y)$. In both MMRM and PMMs,
the posterior distribution of  $(\phiv_y,Y_{\text{m}_1})$ is $\text{pos}_1(\phiv_y,Y_{\text{m}_1}|Y_o,X)$.

We propose the following two-step procedure for imputing  missing data from their posterior predictive distribution in PMMs
\begin{itemize}
\item[1.] Sample $(\phiv_y,Y_{\text{m}_1})$ from the marginal posterior distribution $\text{pos}_1(\phiv_y,Y_{\text{m}_1}|Y_o,X)$ using an algorithm described in Section $2$;
\item[2.] Impute $\yv_{i\text{m}_2}$ from $Q(\yv_{i\text{m}_2}|\vec\yv_{ir_i},\xv_i,\phiv_y)$ for $i=1,\ldots,n$.
\end{itemize}
Because  step $1$ of the  algorithm does not depend on step $2$, it is equivalent to running step $1$
until convergence, and then sampling $\yv_{i\text{m}_2}$'s from $Q(\yv_{i\text{m}_2}|\vec\yv_{ir_i},\xv_i,\phiv_y)$.
The proposed approach is not necessarily a MCMC algorithm. For example, we may run step $1$ via a FDA approach under MAR,
 and then 
run step $2$ to impute $\yv_{i\text{m}_2}$'s in PMMs. However,  the joint distribution of $(\phiv_y,Y_{\text{m}_1},Y_{\text{m}_2})$
will converge to the target distribution \eqref{pospmm}.

Suppose the joint distribution of $\yv_i=(\vec\yv_{is}',\yv_{i\text{m}_2}')'$  is  normal with mean  $(\etav_{is}',\etav_{im}')'$ and covariance  $\Sigma = \begin{bmatrix} \Sigma_{11^s} &   \Sigma_{21^s}' \\
                                             \Sigma_{21^s} & \Sigma_{22^s} \\
                                             \end{bmatrix}$ 
                                           in  pattern $s$. 
The conditional distribution $Q(\yv_{i\text{m}_2}|\vec\yv_{ir_i},\xv_i,\phiv_y)$  of $\yv_{i\text{m}_2}$ given $\vec\yv_{ir_i}$  is  \citep{2013:carpenter}
\begin{equation}\label{cond}
\yv_{i\text{m}_2} |\vec\yv_{is}, r_i=s \sim N[\etav_{im}+ \Sigma_{21^s}\Sigma_{11^s}^{-1}(\vec\yv_{is}-\etav_{is}),
\Sigma_{22^s}-\Sigma_{21^s}\Sigma_{11^s}^{-1}\Sigma_{21^s}'].
\end{equation}
Because all models assume the same observed data distribution and a common covariance matrix $\text{var}(\yv_i)=\Sigma$ across patterns, 
the missing data distribution  $Q(\yv_{i\text{m}_2}|\vec\yv_{ir_i},\xv_i,\phiv_y)$ in these models differs only in $\etav_{im}$.
%Because $Q(\yv_{i\text{m}_2}|\vec\yv_{ir_i},\xv_i,\phiv_y)$ is normal with same covariance matrix in all models,
Therefore, the imputed values in PMMs can be obtained by subtracting the difference in mean $\yv_{i\text{m}_2}$ from the imputed values in MMRM while the imputation for MMRM can be done
using SAS PROC MI. 
All models assume MAR in the control arm, and hence have the same imputed values among control subjects. 

For the purpose of missing data imputation, it is more convenient to express the conditional distribution $Q(\yv_{i\text{m}_2}|\vec\yv_{ir_i},\xv_i,\phiv_y)$ in terms of 
  the LDL decomposition of $\Sigma$.
Suppose  $U$ and $\Lambda$ can be partitioned  as    
                                             $U=\begin{bmatrix} U_{11^s} & \zerov \\
                   U_{21^s} & U_{22^s}
   \end{bmatrix}$, and $\Lambda =\text{diag}(\Lambda_{1^s},\Lambda_{2^s})$
 according to the outcomes before and after dropout in pattern $s$. That is, the dimensions of  $U_{11^s}$,
and $\Lambda_{1^s}$ are all $s\times s$.  Let $L=U^{-1}=  \begin{bmatrix} L_{11^s} & \zerov \\
                   L_{21^s} & L_{22^s}   \end{bmatrix}$.
  Then $\Sigma_{21^s}\Sigma_{11^s}^{-1}=-U_{22^s}^{-1}U_{21^s}=-L_{22^s}U_{21^s}$,
 and $\Sigma_{2.1^s}=\Sigma_{22^s}-\Sigma_{21^s}\Sigma_{11^s}^{-1}\Sigma_{21^s}' = L_{22^s}\Lambda_{2^s}L_{22^s}'$.
In pattern $s=0$, $\Sigma_{2.1^s}=\Sigma_{22^s}=\Sigma$,  $U_{22^s} =U$, and $L_{22^s}=L$. Then \eqref{cond} becomes
\begin{equation}\label{cond2}
\yv_{i\text{m}_2} |\vec\yv_{is}, r_i=s  \sim N[\etav_{im}-L_{22^s}U_{21^s}(\vec\yv_{is}-\etav_{is}), L_{22^s}\Lambda_{2^s}L_{22^s}'].
\end{equation}

 \begin{table}[h]
\footnotesize
\begin{center}
\begin{tabular}{lcccccc} \\\hline 
        &  \multicolumn{2}{c}{Assumption on missing data after dropout}   & & \\ \cline{2-3}
        & Marginal    & $E(y_{ij}|\vec{\yv}_{ij-1},\xv_i)$ & \multicolumn{1}{c}{Conditional mean } &    Mean  \\ 
Method &    mean of $\yv_{i\text{m}_2}$ $^{(a)}$ &  $j>s$ $^{(b)}$  &  of $\yv_{i\text{m}_2}$ given $(\vec{\yv}_{is},\xv_i)$  $^{(c,d)}$  &   Difference $^{(f)}$  \\ \hline
MMRM (MAR)   &  $\muv_{i2^s}+\deltav_{2^s}g$ & $\vartheta_{ij}+ \underline{\delta}_{j}\, g$ &   $\muv^{\text{MAR}}=\muv^{\text{CR}}+L_{22^s}\underline\deltav_{2^s}g$ & $\zerov$  \\  
\multicolumn{3}{l}{Controlled Imputation}\\
\quad J2R &  $\muv_{i2^s}$   & - &    $\muv^{\text{MAR}}-\deltav_{2^s}g$  & $-\deltav_{2^s}g$\\
\quad CIR  &   $\muv_{i2^s}+\delta_s \onev_{p-s}g$ & - &  $\muv^{\text{MAR}}+(\delta_s\onev_{p-s}-\deltav_{2^s})g$ & $(\delta_s\onev_{p-s}-\deltav_{2^s})g$\\
\quad CR &   -   & $\vartheta_{ij}$ &  $\muv^{\text{MAR}}-L_{22^s}\underline\deltav_{2^s}g$   & $-L_{22^s}\underline\deltav_{2^s}g$  \\
\quad ECR  &  -   & $\vartheta_{ij}+ (1-\phi)\underline{\delta}_{j}\, g$  &     $\muv^{\text{MAR}}-\phi L_{22^s}\underline\deltav_{2^s}g$  & $-\phi L_{22^s}\underline\deltav_{2^s}g$ \\
\quad MCR &  - & $\vartheta_{ij}+ d_j \underline\delta_{j}\, g$ &     $\muv^{\text{MAR}}- L_{22^s}\Deltav_{s}^{\text{\tiny MCR}} \,g$ & $- L_{22^s}\Deltav_{s}^{\text{\tiny MCR}} \,g$\\
\multicolumn{3}{l}{Delta-adjusted Imputation}\\
\quad conditional   &  - &  $\vartheta_{ij}+  (\underline\delta_{j}-\Delta_{s_j}^c)\, g$ &  $\muv^{\text{MAR}}- L_{22^s} \Deltav_{s}^c\,g$  & $ - L_{22^s} \Deltav_{s}^c\,g$\\
\quad unconditional &  $\muv_{i2^s}+(\deltav_{2^s}-\Deltav_s^u)g$ & - &  $\muv^{\text{MAR}}-  \Deltav_{s}^u\,g$ & $ -  \Deltav_{s}^u\,g$\\
\hline
 \end{tabular} \caption{\footnotesize
 Missing data distribution in pattern $s$ for various models: 
  $^{(a)}$ the mean of $\vec{\yv}_{is}$ is $\muv_{i1^s}+\deltav_{1^s}g$, and the variance of
  $\yv_i$ is $\Sigma$ in all models. 
 In pattern $s=0$, $\delta_s=0$; $^{(b)}$   $\vartheta_{ij}=\underline\mu_{ij} +\sum_{t=1}^{j-1} \beta_{jt} y_{it}$.
The conditional variance of $y_{ij}$ given $(\vec{\yv}_{ij-1},\xv_i)$ is $\gamma_j^{-1}$.
 $^{(c)}$ the conditional variance  of  $\yv_{i\text{m}_2}$ given $(\vec{\yv}_{is},\xv_i)$
  is $\Sigma_{2.1^s} = L_{22^s}\Lambda_{2^s}L_{22^s}'$; $^{(d)}$ 
$\muv^{\text{CR}} = L_{22^s}(\underline{\muv}_{i2^s}-U_{21^s}\vec{\yv}_{is})$. In pattern $s=0$,   $U_{21^s}\vec{\yv}_{is}=\zerov$; 
$^{(f)}$ Difference in mean of $\yv_{i\text{m}_2}$ between PMM and MMRM. 
}\label{deltat} 
\end{center}
\end{table}

\setlength{\parindent}{3ex}
Below we  briefly describe the assumption on missing data in each model. A summary of the assumptions in all models is provided in Table \ref{deltat}.
The assumption can be formulated based either  on the marginal distribution of $\yv_{i\text{m}_2}$ or on the conditional distribution of $y_{ij}$ given the historical outcome.
The following notations are used.
Suppose subject $i$ is in pattern $r_i=s$ ($s<p$), treatment group $g_i=g$. Let $\ev_i$ be a $p-s$ vector of standard normal random variables.
Let $\mu_{ij}=\sum_{k=0}^{d} \alpha_{jk} x_{ik}$
denote the mean of subject $i$ at  visit $j$ if the subject was on the control treatment,
$\muv_{i1^s}= (\mu_{i1},\ldots,\mu_{is})'$, and $\muv_{i2^s}= (\mu_{i,s+1},\ldots,\mu_{ip})'$. 
Let $\underline\mu_{ij} = \mu_{ij} -\sum_{t=1}^{j-1} \beta_{jt} \mu_{it}=\sum_{k=0}^{d} \underline\alpha_{jk} x_{ik} $,
and $\underline\muv_{i2^s}= (\underline\mu_{i,s+1},\ldots,\underline\mu_{ip})'$. 
Let  $\deltav_{1^s}= (\delta_{1},\ldots,\delta_{s})'$, $\deltav_{2^s}= (\delta_{s+1},\ldots,\delta_{p})'$, 
 $\underline\deltav_{1^s}= (\underline\delta_{1},\ldots,\underline\delta_{s})'$ and 
  $\underline\deltav_{2^s}= (\underline\delta_{s+1},\ldots,\underline\delta_{p})'$. 

\flushleft{\bf a. MMRM (MAR)}

The MMRM  assumes dropouts have the same mean response trajectory as completers with  identical historical outcome and covariates. 
%It implies that the  conditional distribution of $\yv_{i\text{m}_2}$ given  $(\vec{\yv}_{ir_i},\xv_i)$ is the same between dropouts and completers, i.e. 
By \eqref{cond2},
 $\yv_{i\text{m}_2}|\vec{\yv}_{is},\xv_i,r_i=s \sim N(\muv^{\text{\tiny MAR}}, \Sigma_{2.1^s})$,
where $U_{21^s}\vec{\yv}_{is}=\zerov$ at $s=0$, and
$\muv^{\text{\tiny MAR}} = \muv_{i2^s}+\deltav_{2^s}g_i-L_{22^s}U_{21^s}(\vec{\yv}_{is}-\muv_{i1^s}-\deltav_{1^s}g_i) =
L_{22^s}(\underline{\muv}_{i2^s}+\underline\deltav_{2^s}\,g_i-U_{21^s}\vec{\yv}_{is})$.
Thus $\yv_{i\text{m}_2}$ can be generated in matrix form as
\begin{equation}\label{matmmrm}
 \yv_{i\text{m}_2}^{\text{\tiny MAR}} = L_{22^s} (\underline{\muv}_{i2^s}+\underline\deltav_{2^s}\,g_i-U_{21^s}\vec{\yv}_{is} + \Lambda_{2^s}^{1/2} \ev_i),
\end{equation}
or sequentially  from the following regression model
\begin{equation}\label{seqmmrm}
 y_{ij}|\vec{\yv}_{ij-1},\xv_i \sim N\left[\underline\mu_{ij} +\sum_{t=1}^{j-1} \beta_{jt} y_{it}+ \underline{\delta}_{j}\, g_i, \gamma_{j}^{-1}\right]
\text{ for } j>s.
\end{equation}

\setlength{\parindent}{3ex}
Throughout the paper,  the model parameters  in the missing data distribution  (e.g. \eqref{matmmrm} and \eqref{seqmmrm}) are evaluated at the values randomly  drawn from their
posterior distribution, and in SRI (e.g. \eqref{seqmmrm}), the imputed values at previous visits will be used as predictors
for imputing the missing values at the next visit. 
%In SAS,   $\yv_{i\text{m}_2}$'s can be imputed using the MAD-SAS or FDA-SAS algorithms. %The sample SAS code is provided in  the supplementary materials.

\flushleft{\bf b. Control-based Imputation}

The control-based imputation assumes that the statistical behavior of active subjects  after dropout  is similar to that of control subjects, and it reflects
the fact that subjects generally no longer receive the active treatment after dropout. \cite{2013:carpenter} proposed
three control-based PMMs:  jump to reference (J2R), copy increment in reference (CIR), and CR. These PMMs are 
suitable for placebo controlled trials and studies where the control treatment consists of a standard-of-care treatment,
and subjects discontinued from the active arm tend to switch to standard-of-care \citep{ratitch:2013}.
%In these models, the similarity between the active and control subjects is defined respectively in terms of 
%the marginal mean of  $\yv_{i\text{m}_2}$ (J2R),  the mean change %(mean of $(y_{i,s+1}-y_{is},\ldots, y_{ip}-y_{is})'$) 
%from the last observed visit (CIR), or
%the conditional mean of   $\yv_{i\text{m}_2}$  given  $(\vec{\yv}_{is},\xv_i)$  (CR).
 \cite{lu:2014b}  considered an extension of the CR approach,
which uses a sensitivity parameter to capture the gradual departure from the MAR mechanism.
We also suggest a modification of the CR procedure to avoid the possibility that dropouts from the active arm have better mean response trajectory
 than subjects  who remain on the active treatment.

\flushleft{\bf b.1. Jump to Reference (J2R)}

In J2R,  once the active subjects cease the treatment, their mean response jumps to that of the control subjects.
It essentially assumes that immediately upon withdrawal from the active group, all benefit from the treatment is gone \citep{2013:mallinckrodta}.
The mean response is $\muv_{i1^s}+\deltav_{1^s}g_i$ before dropout, and $\muv_{i2^s}$ after dropout.
 The conditional distribution of $\yv_{i\text{m}_2}$ given $(\vec{\yv}_{is},\xv_i)$ is $N(\muv^{\text{\tiny J2R}}, \Sigma_{2.1^s})$,
where $\muv^{\text{\tiny J2R}} =\muv_{i2^s}-L_{22^s}U_{21^s}(\vec{\yv}_{is}-\muv_{i1^s}-\deltav_{1^s}g_i)= 
                      \muv^{\text{\tiny MAR}}-\deltav_{2^s}g_i$ for subjects in pattern $s$.
Thus $\yv_{i\text{m}_2}$ can be generated as
 $$\yv_{i\text{m}_2}^{\text{\tiny J2R}}  =\yv_{i\text{m}_2}^{\text{\tiny MAR}}-\deltav_{2^s}g_i.$$

\flushleft{ \bf b.2. Copy Increment in Reference (CIR)}

The CIR assumes that  the mean profile of active subjects after dropout is  parallel to that of control subjects.
  The mean response is $\muv_{i1^s}+\deltav_{1^s}g_i$ before dropout, and $\muv_{i2^s}+\delta_s \onev_{p-s} g_i$ after dropout,
where $\onev_k$ is a $k\times 1$ vector of ones.
The conditional distribution of $\yv_{i\text{m}_2}$ given $(\vec{\yv}_{is},\xv_i)$ is $N(\muv^{\text{\tiny CIR}}, \Sigma_{2.1^s})$,
where $\muv^{\text{\tiny CIR}} 
 =\muv^{\text{\tiny MAR}}+(\delta_s\onev_{p-s}-\deltav_{2^s})g_i$ for subjects in pattern $s$, and  $\delta_s=0$
at $s=0$  since there is no difference due to treatment in mean baseline response between two arms.
 We can impute
 $\yv_{i\text{m}_2}$ using
$$\yv_{i\text{m}_2}^{\text{\tiny CIR}}  = \yv_{i\text{m}_2}^{\text{\tiny J2R}}+\delta_s\onev_{p-s}g_i=\yv_{i\text{m}_2}^{\text{\tiny MAR}}+(\delta_s\onev_{p-s}-\deltav_{2^s})g_i.$$

\flushleft{\bf b.3  Copy Reference (CR)}\\
In CR, the conditional distribution of $\yv_{i\text{m}_2}$ given  $(\vec{\yv}_{ir_i},\xv_i)$ among  dropouts in the active arm 
is the same as that of control subjects. That is,  $\yv_{i\text{m}_2}|\vec{\yv}_{is},\xv_i,r_i=s \sim N(\muv^{\text{\tiny CR}}, \Sigma_{2.1^s})$,
where $\muv^{\text{\tiny CR}} %=\muv_{i2^s}+\Sigma_{21^s}\Sigma_{11^s}^{-1}( \vec{\yv}_{is} -\muv_{i1^s})
   =  L_{22^s}(\underline{\muv}_{i2^s}-U_{21^s} \vec{\yv}_{is}) = \muv^{\text{\tiny MAR}} - L_{22^s}\underline\deltav_{2^s}\,g_i$. Thus
 $\yv_{i\text{m}_2}$ can be imputed as
  $$\yv_{i\text{m}_2}^{\text{\tiny CR}} =\yv_{i\text{m}_2}^{\text{\tiny MAR}}- L_{22^s}\underline\deltav_{2^s}\,g_i.$$
The missing data 
after dropout in both arms  can also be imputed sequentially from
\begin{equation}\label{zerodose}
 y_{ij}|\vec{\yv}_{ij-1},\xv_i,r_i=s \sim N\left[\underline\mu_{ij} +\sum_{t=1}^{j-1} \beta_{jt} y_{it}, \gamma_{j}^{-1}\right]
\text{ for } j>s,
\end{equation}
which is identical to the zero-dose model of  \cite{little:1996}.
The method is called copy reference possibly because  the missing data  distribution
 is still $N(\muv^{\text{CR}}, \Sigma_{2.1^s})$ if active subjects are assumed to have 
the same mean response profiles  as the reference (i.e. control) subjects both before and after dropout \citep{2013:carpenter}.

\setlength{\parindent}{3ex}
A variant of the CR procedure is implemented in SAS (version 9.4), in which the imputation model is built using only data from the control arm 
\citep{ratitch:2011, ayele:2014}. Although \cite{lu:2014a} showed  that two variants of CR  performed similarly in a simulation study,
the use of only control data may lead to  larger random noise in the imputed outcomes if the sample size is small in the control arm.

\flushleft{\bf b.4. Extension of Copy Reference (ECR)}\\
\cite{lu:2014b}  considered an extension of the zero-dose or CR model, which assumes  
 \begin{equation}\label{missinge}
y_{ij}|\vec{\yv}_{ij-1},\xv_i, r_i=s \sim N\left[\underline{\mu}_{ij} +  (1-\phi)\underline\delta_{j}\, g_i +\sum_{t=1}^{j-1} \beta_{jt}y_{it},
 \gamma_j^{-1}\right]  \text{ for } j>s,
\end{equation}
where $\phi \in [0,1]$  is a pre-specified sensitivity parameter that characterizes
the gradual deviation from the MAR mechanism, with $\phi=0$ corresponding to MAR with the full benefit of the active treatment,
and $\phi=1$ corresponding to the zero-dose model. 
%The conditional distribution of $\yv_{i\text{m}_2}$ given $(\vec{\yv}_{is},\xv_i)$ is $N(\muv^{\text{\tiny ECR}}, \Sigma_{2.1^s})$,
%where $\muv^{\text{\tiny ECR}} = L_{22^s}[\underline{\muv}_{i2^s} +(1-\phi)\underline\deltav_{2^s}g_i-U_{21^s}\vec{\yv}_{is}]= \muv^{\text{\tiny MAR}}-\phi L_{22^s} 
%\underline\deltav_{\,2^s}$.
The dropout missing data  can be imputed from \eqref{missinge} or equivalently from
\begin{equation*}
\yv_{i\text{m}_2}^{\text{\tiny ECR}} =\yv_{i\text{m}_2}^{\text{\tiny MAR}} - \phi L_{22^s} \underline\deltav_{2^s}\,g_i. 
\end{equation*}

\flushleft{\bf b.5. A Modification of Copy Reference (MCR)}\\
Below we illustrate a potential issue with CR using a simple example.  
Suppose there are only two post-baseline visits ($p=2$), and the treatment effects  are positive ($\delta_1>0$ and $\delta_2>0$) at both visits
 (assuming higher scores  represent improvement), but $\underline\delta_2$ is negative.
The mean $y_{i2}$  is $\sum_{k=0}^d \underline\alpha_{2j} x_{ij}+ \beta_{21}\text{E}(y_{i1})+\underline\delta_2$ among active subjects who complete the study,
and  $\sum_{k=0}^d \underline\alpha_{2j} x_{ij}+ \beta_{21}\text{E}(y_{i1})$ among active subjects who discontinue  after the first visit. 
In CR, the dropouts have better mean response at visit $2$ than completers.  

 \setlength{\parindent}{3ex}
We propose a simple modification of the CR procedure. 
Let $d_j=0$ if  $\hat{\underline{\delta}}_j \hat\delta_p\geq 0$, and $1$ otherwise, 
where $\hat{\underline\delta}_j$ and $\hat\delta_p$ are the 
(restricted) maximum likelihood estimates (MLE) from \eqref{mixed3}. Since the true parameters are unknown, $d_j$'s are determined based on the MLE from \eqref{mixed3} prior to  the imputation.
In MCR, the missing value can be imputed sequentially from 
 \begin{equation}\label{mcr}
y_{ij}|\vec{\yv}_{ij-1},\xv_i \sim N\left[\underline{\mu}_{ij} +  d_j \underline\delta_{j}\, g_i +\sum_{t=1}^{j-1} \beta_{jt}y_{it},
 \gamma_j^{-1}\right],
\end{equation}
or equivalently in matrix form from 
\begin{equation}\label{mcrmat}
\yv_{i\text{m}_2}^{\text{\tiny MCR}} =\yv_{i\text{m}_2}^{\text{\tiny MAR}} - L_{22^s} \Deltav_{s}^{\text{\tiny MCR}} \,g_i,
\end{equation} 
where  $\Deltav_s^{\text{\tiny MCR}} =((1-d_{s+1})\underline\delta_{s+1},\ldots,(1-d_p)\underline\delta_p)'$.
The MCR procedure is identical to the CR procedure if $d_1=\ldots=d_p=0$.

\flushleft{\bf c. Delta-adjusted imputation}\\
In the delta-adjusted PMMs, subjects who discontinue from the active treatment will have their unobserved outcome worse
by some pre-specified amount compared with subjects who continue the treatment \citep{ratitch:2013}. The adjustment could be 
applied in either conditional or unconditional ways.  
 
\flushleft{\bf c.1. Conditional delta-adjusted imputation}\\
In the conditional approach, the missing data after dropout in pattern $s$ can be imputed sequentially 
from the following  regression model
 \begin{equation}\label{deltax}
y_{ij}|\vec{\yv}_{ij-1},\xv_i \sim N\left[\underline{\mu}_{ij} +  (\underline\delta_{j}-\Delta_{s_j}^c)\, g_i +\sum_{t=1}^{j-1} \beta_{jt}y_{it},
 \gamma_j^{-1}\right],
\end{equation}
where $\Delta_{s_j}^c$'s are the pre-fixed amount of  adjustment at visit $j>s$ for the active subjects in pattern $s$. 
There are two popular ways to specify $\Delta_{s_j}^c$'s \citep{ratitch:2013, 2013:mallinckrodta}. 
The adjustment can be applied only once at the first visit after dropout (i.e.
 $\Delta_{s_j}^c  =\Delta$ when $j=s+1$, and $0$ if $j>s+1$), or applied to all visits after dropout (i.e.  $\Delta_{s_j}^c  =\Delta$ when $s+1\leq j\leq p$).
The two adjustment strategies correspond respectively to variant-1 and variant-2 described in \cite{ratitch:2013}.

\setlength{\parindent}{3ex}
By \eqref{deltax}, the conditional distribution of $\yv_{i\text{m}_2}$ given $(\vec{\yv}_{is},\xv_i)$ is $N(\muv^{\text{\tiny cDEL}}, \Sigma_{2.1^s})$,
where $\Deltav_s^c=(\Delta_{s_{s+1}}^c,\ldots,\Delta_{s_p}^c)'$ and 
$\muv^{\text{\tiny cDEL}} = L_{22^s}[\underline{\muv}_{i2^s} +(\underline{\deltav}_{2^s}  -\Deltav_{s}^c)\,g_i-U_{21^s}\vec{\yv}_{is}]$.
The missing data after dropout can be imputed from \eqref{deltax} or equivalently from
\begin{equation}\label{delta2}
\yv_{i\text{m}_2}^{\text{\tiny cDEL}} =\yv_{i\text{m}_2}^{\text{\tiny MAR}} - L_{22^s} \Deltav_{s}^c\,g_i. 
\end{equation}

\flushleft{\bf c.2. Unconditional delta-adjusted imputation}\\
The unconditional approach corresponds to variant-3  of \cite{ratitch:2013}.
The adjustment is made by simply subtracting a constant from the MAR-based imputation
\begin{equation}\label{delta4}
\yv_{i\text{m}_2}^{\text{\tiny uDEL}} =\yv_{i\text{m}_2}^{\text{\tiny MAR}} -  \Deltav_{s}^{u} \,g_i, 
\end{equation}
where  $\Deltav_{s}^{u} =  (\Delta_{s_{s+1}}^u,\ldots,\Delta_{s_p}^u)'$ is a vector of pre-specified constants. 
Unlike the conditional approach in which the adjustment at earlier visits will affect subsequent visits, 
the adjustments  at different visits are unrelated in the unconditional approach.
%the amount of adjustment $\Deltav_{s}^{u}$ is fixed across all imputations. But in the conditional approach, 
%the amount of adjustment $L_{22^s} \Deltav_{s}^c$ changes with imputations since $L_{22^s}$'s are random samples from the MCMC algorithm.

\section{A numerical example}\label{numsec}
\setlength{\parindent}{3ex}
We analyze an antidepressant clinical trial reported in \cite{2013:mallinckrodta}. 
The Hamilton 17-item rating scale for depression is collected at baseline 
and weeks 1, 2, 4, 6. The dataset consists of $84$ active subjects, and $88$  placebo subjects.
The dropout rate is $24\%$ ($20/84$) in the active arm, and $26\%$
($23/88$) in the placebo arm. 

 The primary endpoint is the change from baseline in Hamilton depression score, and 
 the explanatory variables  are intercept,  baseline Hamilton score, and  treatment status ($p=4$, $q=3$).
We analyze the data using both frequentist and Bayesian approaches.
In the frequentist approach, the MMRM analysis is fit using SAS PROC MIXED.  It includes
the treatment$\times$visit and baseline$\times$visit interactions as the fixed effects,  and
an unstructured covariance matrix is used to model the within-patient errors. 
Our  model is different from that used in \cite{2013:mallinckrodta} in that we do not include the investigative site as
 a covariate.
In the Bayesian analysis, we consider
three MCMC schemes: the MDA-T algorithm  under the prior  $f(\alphav,\Sigma)\propto |\Sigma|^{-(p+3)/2}$ (i.e. $\nu_0=2$), and 
the MDA-SAS and FDA-SAS algorithms under Jeffrey's prior $f(\alphav_w,\Sigma_w)\propto |\Sigma_w|^{-(p+q)/2}$.
By Lemma \ref{sasalg}, the three MCMC schemes yield the same posterior distribution of $(\alphav,\Sigma)$.
The latter two analyses are conducted using SAS PROC MI.
As shown in Table \ref{examres1}, the treatment effects from the three Bayesian analyses and the likelihood-based analysis
are similar since the prior is non-informative.

\begin{table}[h]
\begin{center}
\scriptsize
%\footnotesize
\begin{tabular}{lcccccccccccccccccc} \\\hline 
&& \multicolumn{3}{c}{Bayesian analysis$^a$}\\\cline{3-5}
Week & Proc Mixed &  MDA-T & MDA-SAS & FDA-SAS$^b$  \\\hline
1 & $0.092\,[-1.256, 1.439]$ & $0.092 \, [-1.251, 1.436]$ & $0.091\,[-1.267,   1.446]$ & $0.096\,[-1.245, 1.462]$\\
2  & $-1.403\,[-3.228,0.422]$& $-1.404\, [-3.230, 0.434]$ & $-1.402\,[-3.221, 0.430]$& $-1.395\,[-3.200,   0.440]$\\
4  & $-2.225\,[-4.201,-0.248]$& $-2.219\,[-4.209, -0.235]$ & $-2.229\, [-4.213, -0.233]$& $-2.220\,[-4.191,  -0.251]$\\
6 & $-2.802\,[-5.008, -0.596]$ & $-2.793\,[-5.004, -0.587]$ & $-2.803\,[-5.016, -0.586]$& $-2.795\,[-5.012, -0.557]$\\
\hline
 \end{tabular} \caption{\footnotesize Estimated treatment effect [$95\%$ confidence or credible interval]  in an antidepressant trial:
  $^a$ posterior mean and quantile-based credible interval are evaluated based on $40,000$ MCMC samples
 collected from every $100$-th iteration  after a ``burn-in'' period of $10,000$ iterations; $^b$
 In FDA-SAS, Jeffrey's prior can be specified with  the    
 PRIOR=INPUT= option (adjusted for $1$ df in the prior) or PRIOR=JEFFREYS option, and both options produce identical output. 
 }\label{examres1}
\end{center}
\end{table}

We also analyze the data using various PMMs, where $m=10,000$ posterior samples are collected from the MDA-SAS algorithm, and 
the complete datasets are imputed using the strategy in Section $3$.  Table \ref{examres2} displays the treatment effect estimate
$\pm$ standard error (SE) in MMRM and various PMMs. Under MAR,  the MI-based  and likelihood-based analyses
yield very close results. At week $1$, there is no missing data, and the treatment effect estimates are identical in all approaches.
In weeks $2$, $4$, and $6$, the PMMs generally yield smaller treatment effect estimate than the MMRM.
The MCR and CR produce the same result since $\hat{\underline\delta}_j\hat\delta_4\geq 0$ at $j\geq 2$, and $y_{i1}$'s are
observed for all subjects although $\hat{\underline\delta}_1\hat\delta_4<0$. The sample SAS code is provided in the supplementary materials.

\begin{table}[h]
\begin{center}
\footnotesize
\begin{tabular}{lcccccccccccccccccc} \\\hline 
Method & Week $1$ & Week $2$ & Week $4$ & Week $6$ \\\hline
\multicolumn{2}{l}{\bf Missing at random }\\
\quad MMRM (ML) & $ 0.092\pm 0.683$ & $-1.403\pm0.924$ & $-2.225\pm 1.001$ & $-2.802\pm1.116$\\
\quad MMRM (MI)  &$ 0.092\pm 0.683$&	$-1.401\pm 0.925$&	$-2.224\pm 1.001$&	$-2.806\pm 1.118$\\ 
\multicolumn{2}{l}{\bf Missing not at random } \\
\multicolumn{2}{l}{\it Control-based imputation} \\
\qquad J2R & $ 0.092\pm 0.683$&	$-1.303\pm 0.927$&	$-1.927\pm 1.004$&	$-2.126\pm 1.130$\\
\qquad CIR & $ 0.092\pm 0.683$&	$-1.296\pm 0.926$&	$-2.009\pm 1.001$&	$-2.451\pm 1.109$\\
\qquad CR  & $ 0.092\pm 0.683$&	$-1.297\pm 0.926$&	$-1.975\pm 1.001$&	$-2.372\pm 1.109$\\
\qquad ECR: $\phi=0.5$ & $ 0.092\pm 0.683$&	$-1.349\pm 0.925$&	$-2.100\pm 0.999$&	$-2.589\pm 1.109$\\
\qquad MCR & $ 0.092\pm 0.683$&	$-1.297\pm 0.926$&	$-1.975\pm 1.001$&	$-2.372\pm 1.109$\\
% LMCF & $ 0.092\pm 0.683$&	$-1.303\pm 0.931$&	$-2.015\pm 1.005$&	$-2.496\pm 1.143$\\
\multicolumn{2}{l}{\it Conditional delta-adjusted imputation} \\
 \quad $\Deltav_s^c=(-4,0,\dots,0)'$ & $ 0.092\pm 0.683$&	$-1.122\pm 0.938$&	$-1.800\pm 1.009$&	$-2.020\pm 1.141$\\
  \quad $\Deltav_s^c=(-2,\dots,-2)'$ & $ 0.092\pm 0.683$&	$-1.261\pm 0.930$&	$-1.873\pm 1.010$&	$-2.047\pm 1.139$\\
\multicolumn{2}{l}{\it Unconditional delta-adjusted imputation} \\
  \quad $\Deltav_s^u=(-3,\ldots,-3)'$ & $ 0.092\pm 0.683$&  $-1.192\pm  0.934$ & $-1.826\pm  1.010$ & $-2.082\pm 1.136$\\
\hline
 \end{tabular} \caption{\footnotesize Estimated treatment effect $\pm$ SE by visit in MMRM and various PMMs. PMMs are implemented via MI
 based on  $m=10,000$ imputations. Posterior samples are 
 collected from every $100$-th iteration  after a ``burn-in'' period of $10,000$ iterations using the MDA-SAS algorithm.
 }\label{examres2}
\end{center}
\end{table}

\section{Comparison with several existing methods}
This section reviews several existing imputation methods. These methods
 generally 
produce similar results to the algorithm proposed in section $3$ if they make the same assumptions on the observed data distribution (i.e. same observed data likelihood), 
and  the number of imputations is large enough to stabilize the result. The difference in results is usually small, and it arises because of the use of different priors, and imputation variability
due to the use of a finite number of imputations.

The main advantages of the proposed algorithm are  (1) it  sufficiently uses the existing functions of SAS PROC MI, and the SAS code is simpler and easier to maintain and/or modify, 
and (2) the algorithm reaches stationarity quickly particularly if one chooses the MDA-SAS approach for imputation in MMRM. It is also more convenient to use the proposed 
method  to compare different PMMs in sensitivity analysis. One can simply save the imputed datasets for MMRM, and the posterior samples of $\phiv_y$, and then use the method
described in Section $3$ to derive the imputed dataset for each PMM.

\setlength{\parindent}{3ex}
\subsection{Sequential regression imputation (SRI)} 
The SRI approach \citep{little:1996, ratitch:2013, lu:2014a} is a popular imputation method 
for  monotone data. It can  be viewed as a special case of the MDA-T algorithm.
Under monotone missingness, the MDA-T algorithm involves only the P-step, and reaches stationary in one step. 
When $A=\zerov$, $M=\zerov$ and $p(\alphav,\Sigma)\propto |\Sigma|^{-(\nu_0+p+1)/2}$, the posterior
 distribution \eqref{poslimit0} in Appendix \ref{mcmctang} can be  expressed as \citep{2016:tang}
\begin{equation}\label{seqimp}
 \gamma_j| Y_{\text{m}_1},Y_{o} \sim \chi_{n_j+\nu_0+j-q-p}^2/\hat{S}_j \text{ and } 
         \thetav_j|\gamma_j, Y_{\text{m}_1},Y_{o} \sim N[\hat\thetav_j,(\gamma_j\sum_{i=1}^{n_j}\zv_{i,j-1}'\zv_{i,j-1})^{-1}],
\end{equation}
where $\hat\thetav_j =(\sum_{i=1}^{n_j}\zv_{i,j-1}'\zv_{i,j-1})^{-1} (\sum_{i=1}^{n_j}\zv_{i,j-1}'y_{ij})$
and  $\hat{S}_j =\sum_{i=1}^{n_j}(y_{ij}-\zv_{i,j-1}'\hat\thetav_j)^2$. The missing data after dropout can be sequentially
imputed  from \eqref{seqmmrm}, \eqref{zerodose}, \eqref{missinge}, \eqref{mcr}, \eqref{deltax} respectively 
for MMRM, CR, ECR, MCR, and conditional delta-adjusted PMM. 

In the literature, the SRI approach is mainly applied to models with simple conditional distribution of $y_{ij}$ given $\vec{\yv}_{ij-1}$. But
it is also suitable for  complicate PMMs such as J2R and CIR using the method described in Section \ref{meandiffpmm}.

The SRI approach is available in SAS (version 9.4) PROC MI for imputations under MAR, and the conditional delta-adjusted imputations (using MNAR and MONOTONE statements).
In SAS, the posterior distribution of  $\gamma_j\sim \chi_{n_j-q-(j-1)}^2/\hat{S}_j$ is slightly different from \eqref{seqimp}
due to the use of a different prior.

\subsection{\cite{ratitch:2013} approach}
\cite{ratitch:2013} described a procedure for data with intermittent missing values. It firstly uses \cite{schafer:1997}  MDA algorithm
implemented in SAS to impute the intermittent missing data ($Y_{\text{m}1}$) $m$ times. The SRI approach is then used to impute $Y_{\text{m}2}$ 
for each imputed monotone dataset. The sampling schemes are different  in \cite{ratitch:2013} method and the proposed algorithm (assume MDA algorithm is used)
\begin{itemize}
\item Ratitch et al scheme:  Iterate between $\phiv_x, \phiv_y | X, Y_{o}, Y_{\text{m}1}$ and  $Y_{\text{m}1}| \phiv_x, \phiv_y, X, Y_{o}$ until convergence.
                              Sample $\phiv_y^*|X, Y_{\text{m}1}, Y_{o}$, and $Y_{\text{m}2}|\phiv_y^*,X, Y_{\text{m}1}, Y_{o}$ after convergence.
\item Proposed  scheme: Iterate between $\phiv_x, \phiv_y |X, Y_{o}, Y_{\text{m}1}$ and  $Y_{\text{m}1}| \phiv_x, \phiv_y, X, Y_{o}$ until convergence.
            Sample  $Y_{\text{m}2}|\phiv_y,X, Y_{\text{m}1}, Y_{o}$ after convergence.
\end{itemize}
Compared to \cite{ratitch:2013} method,  the proposed scheme avoids one additional step in sampling $\phiv_y^*$, and
it can save computational time  particularly if the number of imputations $m$ is large, or if one wants to compare different PMMs in sensitivity analysis.

Similarly to SRI, \cite{ratitch:2013} method is developed for the CR and  conditional delta-adjusted imputation, but it can be modified to handle 
J2R and CIR.

\subsection{Macro based on SAS Proc MCMC}
\cite{2013:mallinckrodta} developed a SAS package for missing data imputation in PMMs, and it
 is freely available at  {\it http://www.missingdata.org.uk}.  The package is  based on SAS PROC MCMC, in which $(\alphav, \Sigma)$ is sampled using Metropolis-type algorithms. 
The convergence of the Metropolis-type algorithm can be slow particularly if the dimension of parameter space is large (e.g. when the number of post-baseline visits $p$ is large).
The real clinical data are usually monotone or approximately monotone, and the MDA algorithm generally converges much more quickly with smaller autocorrelation between posterior samples than the Metropolis-type algorithm. 
For the antidepressant trial analyzed in Section \ref{numsec}, the MDA algorithm converges  within $100$ iterations, and the lag-$1$ autocorrelation is  close to $0$ for all model parameters
\citep{2016:tang}. Furthermore, our SAS code is much simpler, and  runs much faster than  \cite{2013:mallinckrodta} macro for the same number of MCMC iterations.
But \cite{2013:mallinckrodta} macro can handle more complex MMRM (e.g. covariance matrix heterogeneity).

\section{Discussion}
\setlength{\parindent}{3ex}
The PMMs have been widely used as sensitivity analysis of longitudinal outcomes with non-ignorable missing data. 
We describe a novel  approach for
 missing value imputations in  the MMRM, delta-adjusted PMMs and control-based  PMMs.    The imputed values in PMMs can be obtained from that under MAR by 
subtracting the mean difference in their posterior predictive distributions, 
which is a function of the posterior samples of the MMRM model parameters $\phiv_y$. We have focused on the control-based 
 and delta-adjustment PMMs. However, the imputation algorithm works
for any PMMs that assume the same observed data distribution as  MMRM.

For CR, it is possible to impute missing data in an alternative way using SAS PROC MI. One may firstly impute the intermittent missing data $\yv_{i\text{m}_1}$'s 
under MAR. The dropout missing data $\yv_{i\text{m}_2}$'s
 can be imputed by setting the treatment status as placebo for dropouts in both arms.
%The code would be similar to the sample code provided in the supplementary materials. 

In \eqref{mixed3},  the covariance matrix is assumed to be homogeneous across all subjects, and 
there is a covariate $\times$ visit interaction for each covariate,
so that  the MCMC sampling can be easily implemented using SAS PROC MI.
For more complex MMRMs, one may use  \cite{2016:tang} MDA algorithm or 
\cite{2013:mallinckrodta} SAS macro.
 
There is considerable debate regarding the appropriateness of using Rubin's variance estimator when the data imputation and analysis models are
uncongenial \citep{1994:meng}.
In control-based PMMs, Rubin's variance estimator overestimates the variance of the estimated treatment
effect  \citep{lu:2014a, ayele:2014}. Intuitively, this is because the data are imputed on an as-treated basis \citep{little:1996}, but
analyzed under the ITT principle. In a companion paper, we show that  Rubin's variance estimator  is approximately
unbiased in delta-adjusted PMMs, and this provides theoretical support for the use of delta-adjusted PMMs as MNAR sensitivity analysis in clinical trials.
%In that work, we also describe how to  find analytically the tipping point $\Delta$ value at which the significance of the primary analysis is lost in delta-adjusted PMMs.

Imputations under MNAR  based on the fully conditional specification (FCS) method  are implemented in SAS (version 9.4).  
The FCS approach involves
specifying the  conditional distribution for each incomplete variable given the other variables, and iterating the imputations  on a variable-by-variable basis 
until convergence \citep{Buuren:2007}. 
%The  FCS approach is attractive under MAR  particularly
%if it is hard to specify the joint distribution of those variables with missing data (e.g. some variables are continuous, and some are discrete). 
It can be challenging to find the stationary distribution of the missing data in the FCS approach under MNAR, but it is different from that 
 in the  MCMC method since
the intermittent missing data are MAR in the MCMC algorithms (see Section $3$), and MNAR in the FCS approach
(this can  easily be seen in the special case where all outcomes are observed except few subjects miss the first visit). 
 Further research shall be done to understand the FCS imputation under MNAR.

\bigskip
\begin{center}
{\bf \Large Acknowledgments}
\end{center}
We would like to thank the associate editor and referees for their  constructive comments that greatly help to improve the quality of the article.

\bigskip
\begin{center}
{\large\bf SUPPLEMENTARY MATERIAL}
\end{center}
Sample SAS code for the analysis of the antidepressant trial is provided in the supplementary materials. The raw dataset is freely available  at {\it www.missingdata.org.uk}.

\appendix
\section{Appendix}
\subsection{MDA and FDA algorithms for MMRM}\label{mcmctang}
\setlength{\parindent}{3ex}
\cite{tang:2015, 2016:tang} developed the MDA algorithm (called MDA-T) for MMRM. Compared to  \cite{schafer:1997} MDA algorithm 
for incomplete multivariate normal data, the new MDA method
 allows the  use of both non-informative and informative priors, and  greatly reduces the amount of matrix inversion in imputing 
$(\yv_{i\text{m}_1}, \yv_{i\text{m}_2})$'s from their posterior predictive distribution. 
In addition,  the new MDA algorithm can handle more complex assumptions on the mean and covariance of $\yv_i$'s.

In the prior, $(\thetav_j,\gamma_j)$'s follow independent normal-gamma distributions  \citep{tang:2015} 
 \begin{equation}\label{priortheta}
 f(\thetav_j, \gamma_j)   \propto  \gamma_j^{ \frac{\nu_0+2j+r-p-3}{2}} 
      \exp\left[-\frac{\gamma_j}{2} \tilde{\thetav}_j'(\tilde{A}_j+B_j)\tilde{\thetav}_j\right],
\end{equation}
where $\tilde{\thetav}_j=(-\thetav_j',1)'$, $\tilde{A}_j$ is the  leading $(q+j)\times (q+j)$ submatrix of the $(q+p)\times (q+p)$ matrix
        $\tilde{A}=\begin{bmatrix} \zerov & \zerov \\
                                        \zerov  & A\\
                                        \end{bmatrix}$,
                                        $I_q$ is the $q\times q$ identity matrix, 
$\ddot\alphav_j= (I_q,\alphav_1,\ldots,\alphav_j)$ and $B_j=\ddot\alphav_j' M \ddot\alphav_j$.

The likelihood function for the augmented monotone data is given  by
\begin{equation}\label{likelihood1}
\prod_{i=1}^n  f(\yv_{ir_i}|\xv_i,\phiv_y)
=\prod_{j=1}^p \gamma_j^{\frac{n_j}{2}}\exp\left[-\frac{\gamma_j}{2}\tilde{\thetav}_j'  (\sum_{i=1}^{n_j} \zv_{ij}\zv_{ij}')  \tilde{\thetav}_j\right]. 
\end{equation}
Combining \eqref{priortheta} and \eqref{likelihood1} yields the posterior distribution for$(\thetav_j, \gamma_j)$'s, which is normal-gamma, and 
can be generated using the  methods of  \cite{2016:tang,tang:2015}
\begin{equation}\label{poslimit0}  
 f(\thetav_j, \gamma_j|Y_{\text{m}_1},Y_{o},X)   
\propto   \gamma_j^{ \frac{n_j+\nu_0+2j+r-p-3}{2}}       \exp\left[-\frac{\gamma_j}{2} \tilde{\thetav}_j' D_j \tilde{\thetav}_j\right],
\end{equation}
where $D_j=\tilde{A}_j+B_j+\sum_{i=1}^{n_j} \zv_{ij}\zv_{ij}'$. \cite{2016:tang} derived the posterior distribution of $\yv_{i\text{m}_1}$ 
\begin{equation}\label{missingint}
f(\yv_{i\text{m}_1}|\yv_{io},\phiv_y) 
\propto \prod_{j=h}^{r_i} \exp\left[-\frac{ \gamma_j(\tilde{U}_{jm}'\yv_{i\text{m}_1}  - e_{ij})^2}{2}\right] 
\propto N(\hat{\mu}_{y_{\text{m}_1}}, \hat{V}_{y_{\text{m}1}}),
\end{equation}
where $h$ ($h<r_i$) is the index of the first missing observation for subject $i$, 
 $(\tilde{U}_{jm},  \tilde{U}_{jo})$ is a partition of the  $ (r_i-h+1)\times 1$  vector
$\tilde{U}_j=(-\beta_{jh},\ldots, -\beta_{j,j-1},1,0,\ldots,0)'$ according to the missing ($\yv_{i\text{m}_1}$) and observed ($\yv_{io_h}$) parts of $(y_{ih},\ldots,y_{ir_i})'$ for $j\geq h$, 
$e_{ij}=    \underline\alphav_j'\xv_i +  \sum_{t=1}^{h-1} \beta_{jt}y_{it}- \tilde{U}_{jo}'\yv_{io_h}$ ($\tilde{U}_{jm}'\yv_{i\text{m}_1}  -e_{ij}=y_{ij}-\underline\alphav_j'\xv_i -\sum_{t=1}^{j-1} \beta_{jt}y_{it}$),
$\hat{V}_{y_{\text{m}1}}= (\sum_{j=h}^{r_i} \gamma_j\tilde{U}_{jm}\tilde{U}_{jm}')^{-1}$ and
$\hat{\mu}_{y_{\text{m}_1}}=\hat{V}_{y_{\text{m}1}} \sum_{j=h}^{r_i} \gamma_j e_{ij} \tilde{U}_{jm}$.

The  MDA-T algorithm repeats  the following I- and P- steps until  convergence
  \begin{itemize}%[leftmargin=.6in]  
  \item[P:] Draw $(\thetav_j,\gamma_j)$ from  the posterior distribution \eqref{poslimit0} for $j=1,\ldots,p$;
  \item[I:] Impute $\yv_{i\text{m}_1}$'s from \eqref{missingint} for subjects with intermittent missing data.
  \end{itemize}

Similarly, a FDA algorithm can be developed on basis of
the posterior distribution of $(\thetav_j, \gamma_j)$ given the augmented full data $Y_f=\{Y_o,Y_{\text{m}_1}, Y_{\text{m}_2}\}$
 \begin{eqnarray}\label{posterior2}
     \begin{aligned}
 f(\thetav_j, \gamma_j|Y_f,X)  
\propto \gamma_j^{ \frac{n+\nu_0+2j+r-p-3}{2}} 
      \exp\left[-\frac{\gamma_j}{2} \tilde{\thetav}_j'(\tilde{A}_j+B_j+\sum_{i=1}^{n} \zv_{ij}\zv_{ij}')\tilde{\thetav}_j\right].
      \end{aligned}
\end{eqnarray} 
The algorithm (called FDA-T)   repeats the following steps until  convergence.
  \begin{itemize}%[leftmargin=.6in]  
  \item[P:] Draw $(\thetav_j,\gamma_j)$ from  its posterior distribution \eqref{posterior2} for $j=1,\ldots,p$;
  \item[I:] Impute $\yv_{i\text{m}_1}$'s from \eqref{missingint}, and $\yv_{i\text{m}_2}$'s from $f(\yv_{i\text{m}_2}|\vec\yv_{ir_i},\xv_i,\phiv_y)$.
 \end{itemize}

%\cite{2016:tang} demonstrated that the limiting  distribution of $\phiv$ is the same in the MDA-T and FDA-T algorithms.

\subsection{Proof of Lemma \ref{sasalg}}\label{proof}

\begin{proof} 
Let $\bar{M} =\begin{bmatrix}  0 & \zerov'\\ \zerov & M^* \\\end{bmatrix}$ and $\underline{M}=\begin{bmatrix}  m_{11} & M_{12}\\ M_{21} & \underline{M}^* \\\end{bmatrix}$,
where $\underline{M}^*=M_{21}M_{12}/m_{11}$ if $m_{11}>0$, and $\underline{M}^*=\zerov$ if $m_{11}=0$. Then $\bar{M}+\underline{M}=M$.
Let $\tilde{j}=q-1+j$.
By  Lemma $2$ of  \cite{tang:2015},  in the prior, $(\thetav_{j}^*, \gamma_{j}^*)$'s are independent, and the  distribution of $(\thetav_{\tilde{j}}^*, \gamma_{\tilde{j}}^*)$ is 
\begin{equation*}
f(\thetav_{\tilde{j}}^*, \gamma_{\tilde{j}}^*) \propto {\gamma_{\tilde{j}}^*}^{\frac{\nu_w+2(q-1+j)+r_w-(q-1+p)-3}{2}} \exp\left[-\frac{\gamma_{\tilde{j}}^*}{2} \tilde{\thetav}_{\tilde{j}}^{*'}(\tilde{A}_{w_j}+B_{w_j})\tilde{\thetav}_{\tilde{j}}^*\right] 
\end{equation*}
where  $\tilde{A}_{w}=\begin{bmatrix} 0 & \zerov_{\tilde{p}}' \\ \zerov_{\tilde{p}} & A_{w} \\\end{bmatrix} 
  = \begin{bmatrix} \bar{M} & \bar{M} {\alphav_0}' \\ \alphav_0\bar{M} & A+ \alphav_0\bar{M}\alphav_0'\\ \end{bmatrix}$, $\tilde{\thetav}_{\tilde{j}}^*=(-\thetav_{\tilde{j}}^{*'},1)'$,
$u_{w}=(1, \alphav_{w0}')'$, $B_{w}=m_{11}u_{w}'u_{w}$,
$\tilde{A}_{w_j}$ and $B_{w_j}$ denote respectively the leading $(q+j)\times (q+j)$ submatrix of $A_w$ and $B_w$,
$r_w=1$ if $m_{11}>0$, and $r_w=0$ if $m_{11}=0$.
 A little algebra
shows that $\tilde{A}_{w_j}=\ddot\alphav_j'\bar{M}\ddot\alphav_j+\tilde{A}_j$, $B_{w_j}=\ddot\alphav_j'\underline{M}\ddot\alphav_j$ and $\tilde{A}_{w_j}+B_{w_j}=B_j+\tilde{A}_j$. Thus 
\begin{equation}\label{priornew}
f(\thetav_{\tilde{j}}^*, \gamma_{\tilde{j}}^*) \propto {\gamma_{\tilde{j}}^*}^{\frac{\nu_0+2j+r-p-3}{2}} \exp\left[-\frac{\gamma_{\tilde{j}}^*}{2} \tilde{\thetav}_{\tilde{j}}^{*'}(\tilde{A}_j+B_j)\tilde{\thetav}_{\tilde{j}}^*\right],
\end{equation}
and  $(\thetav_{\tilde{j}}^*, \gamma_{\tilde{j}}^*)$ has the same prior distribution as  $(\thetav_{j}, \gamma_{j})$ defined in \eqref{priortheta}.

Since $\phiv_x$ and $\phiv_y$ are independent in the prior, and
the likelihood function can be factored as the product of 
$ \prod_{i=1}^n f(\tilde\xv_{i} | \phiv_x)$ and $ \prod_{i=1}^n f(\yv_{io}| \xv_i,\phiv_y)$, 
$\phiv_x$ and $\phiv_y$ are independent in the posterior distribution.
The posterior distribution of $\phiv_y$ is
$f( \phiv_y|Y_{o},X)\propto  \prod_{i=1}^n f(\yv_{io}| \xv_i,\phiv_y)
     \prod_{j=1}^p f(\thetav_{\tilde{j}}^*, \gamma_{\tilde{j}}^*)$, which is identical to that from any valid MCMC  algorithm for \eqref{mixed3}.

Since $(\alphav_x,\Sigma_{xx})$ is a function of $\phiv_x$, and $(\alphav,\Sigma)$ is a function of $\phiv_y$, 
$(\alphav_x,\Sigma_{xx})$ and $(\alphav,\Sigma)$ are independent in both the prior and posterior distributions, and the posterior distribution of $(\alphav,\Sigma)$ is identical 
 that from any valid MCMC  algorithm for \eqref{mixed3}.
\end{proof}

\subsection{MCMC algorithms in SAS}\label{mcmcsas}

The   MDA-SAS algorithm  repeats the following steps until the algorithm converges.
%\vspace{-.5in}
 \begin{itemize}%[leftmargin=.6in]
 \item[P1:] Sample  $(\thetav_{j}^*,\gamma_{j}^*)$'s from  $f(\thetav_{j}^*,\gamma_{j}^*|X)$ for $j=1,\ldots,q-1$.
 \item[P2:] Draw $(\thetav_{q-1+j}^*,\gamma_{q-1+j}^*)$'s from $f(\thetav_{q-1+j}^*,\gamma_{q-1+j}^*|Y_{\text{m}_1},Y_{o},X)$ for $j=1,\ldots,p$.
 \item[I:] Impute missing data $\yv_{i\text{m}_1}$'s from $f(\yv_{i\text{m}_1}|\yv_{io},\xv_i,\phiv_y)$.
 \end{itemize}
The P2 and I steps in the MDA-SAS algorithm are equivalent to the P and I steps in the MDA-T algorithm, and are unrelated to
the P1-step. This explains the equivalence of the MDA-T and MDA-SAS algorithms.

The FDA-SAS algorithm  iterates between the following  steps  until  convergence.
%\vspace{-.3in}          
  \begin{itemize}%[leftmargin=.6in]
 \item[I:] Impute missing data $\yv_{i\text{m}_1}$'s and $\yv_{i\text{m}_2}$'s;
 \item[P:] Sample $(\alphav_w,\Sigma_w)$  from their posterior distribution
 \end{itemize}
  {\small
  \begin{eqnarray}\label{posfull}
 \begin{aligned}
  \Sigma_w|Y_f,X & \sim \mathcal{W}^{-1}\left(\sum_{i=1}^n (\wv_i-\bar{\wv})^{\otimes2} +A_w+\frac{n m_{11}}{n+m_{11}} (\bar{\wv}-\alphav_{w0})^{\otimes 2},f\right),\\
  \alphav_{w}|\Sigma_w, Y_f,X & \sim N\left( \frac{1}{n+m_{11}}(n\bar{\wv} +m_{11}\alphav_{w0}),  \frac{1}{n+m_{11}}\Sigma_w\right), \\
  \end{aligned}
  \end{eqnarray}
  }
  where $A^{\otimes2}=AA'$, $\bar{\wv} =n^{-1} \sum_{i=1}^n \wv_i$,  $f=n+\nu_w-1$ if $m_{11}=0$ and $f=n+\nu_w$ if $m_{11}>0$.
The P-steps of the FDA-SAS and FDA-T algorithms seem quite different. If $(\alphav_w,\Sigma_w)$ is distributed as \eqref{posfull}, 
$(\thetav_{q-1+j}^*, \gamma_{q-1+j}^*)$ has in fact the same posterior distribution as  $(\thetav_j,\gamma_j)$ defined \eqref{posterior2}, and this can be proved 
using the fact that the random Wishart matrix can be expressed as a function of independent normal-gamma random variables \citep{tang:2015}.

\cite{schafer:1997} obtained \eqref{posfull} under the normal-inverse-Wishart prior (i.e. $m_{11}\neq 0$). The proof will be similar 
  when  $m_{11}=0$,  at which \eqref{posfull} reduces to
  \begin{eqnarray*}
 \begin{aligned}
  \Sigma_w|Y_f,X & \sim \mathcal{W}^{-1}\left(\sum_{i=1}^n (\wv_i-\bar{\wv})^{\otimes2} +A_w,n+\nu_w-1\right),\\
  \alphav_{w}|\Sigma_w, Y_f,X & \sim N\left(\bar{\wv},  n^{-1}\Sigma_w\right). \\
  \end{aligned}
  \end{eqnarray*}

 SAS  (version 9.4) PROC MI uses  wrong degrees of freedom (df) $f=n+\nu_w$  when $m_{11}=0$ (i.e. $f(\alphav_w|\Sigma_w)\propto \text{constant}$), 
and one specifies the prior  with the PRIOR =INPUT= option. 
 One may reduce the df in the prior  by $1$ (i.e. from $\nu_w$ to $\nu_w-1$) in order 
 to get the right posterior distribution.
The df is right in SAS PROC MI if $m_{11}>0$, or if one specifies Jeffrey's prior using  the PRIOR=Jeffreys  option ($m_{11}=0$).

%\section{BibTeX}
\bibliographystyle{Chicago} 
\bibliography{missingsas_pmm} 

\begin{thebibliography}{}

\bibitem[\protect\citeauthoryear{Ayele, Lipkovich, Molenberghs, and
  Mallinckrodt}{Ayele et~al.}{2014}]{ayele:2014}
Ayele, B.~T., I.~Lipkovich, G.~Molenberghs, and C.~H. Mallinckrodt (2014).
\newblock A multiple-imputation-based approach to sensitivity analyses and
  effectiveness assessments in longitudinal clinical trials.
\newblock {\em Journal of Biopharmaceutical Statistics\/}~{\em 24(2)}, 211 --
  28.

\bibitem[\protect\citeauthoryear{Carpenter, Roger, and Kenward}{Carpenter
  et~al.}{2013}]{2013:carpenter}
Carpenter, J., J.~Roger, and M.~Kenward (2013).
\newblock Analysis of longitudinal trials with protocol deviation: a framework
  for relevant, accessible assumptions, and inference via multiple imputation.
\newblock {\em Journal of Biopharmaceutical Statistics\/}~{\em 23}, 1352 -- 71.

\bibitem[\protect\citeauthoryear{{CHMP}}{{CHMP}}{2010}]{chmp:2010}
{CHMP} (2010).
\newblock {\em EMA Guideline on Missing data in Confirmatory Clinical Trials
  (EMA/CPMP/EWP/1776/99)}.
\newblock London: CHAMP.

\bibitem[\protect\citeauthoryear{{ICH E9}}{{ICH E9}}{1999}]{ICHE9}
{ICH E9} (1999).
\newblock Statistical principles for clinical trials: Ich harmonized tripartite
  guideline.
\newblock {\em Statistics in Medicine\/}~{\em 18}, 1905 -- 42.

\bibitem[\protect\citeauthoryear{Little and Yau}{Little and
  Yau}{1996}]{little:1996}
Little, R. and L.~Yau (1996).
\newblock Intent-to-treat analysis for longitudinal studies with drop-outs.
\newblock {\em Biometrics\/}~{\em 52}, 1324 -- 33.

\bibitem[\protect\citeauthoryear{Liu}{Liu}{1994}]{liu2:1994}
Liu, J.~S. (1994).
\newblock The collapsed {G}ibbs sampler in {B}ayesian computations with
  applications to a gene regulation problem.
\newblock {\em Journal of the American Statistical Association\/}~{\em 89}, 958
  -- 66.

\bibitem[\protect\citeauthoryear{Lu}{Lu}{2014a}]{lu:2014a}
Lu, K. (2014a).
\newblock An analytic method for the placebo-based pattern-mixture model.
\newblock {\em Statistics in Medicine\/}~{\em 33}, 1134--45.

\bibitem[\protect\citeauthoryear{Lu}{Lu}{2014b}]{lu:2014b}
Lu, K. (2014b).
\newblock An extension of the placebo-based pattern-mixture model.
\newblock {\em Pharmaceutical Statistics\/}~{\em 13}, 103--9.

\bibitem[\protect\citeauthoryear{Mallinckrodt, Roger, Chuang-stein,
  Molenberghs, Lane, O'kelly, Ratitch, Xu, Gilbert, Mehrotrak, Wolfinger, and
  Thijs}{Mallinckrodt et~al.}{2013}]{2013:mallinckrodta}
Mallinckrodt, C., J.~Roger, C.~Chuang-stein, G.~Molenberghs, P.~W. Lane,
  M.~O'kelly, B.~Ratitch, L.~Xu, S.~Gilbert, D.~V. Mehrotrak, R.~Wolfinger, and
  H.~Thijs (2013).
\newblock Missing data: Turning guidance into action.
\newblock {\em Statistics in Biopharmaceutical Research\/}~{\em 5}, 369 -- 82.

\bibitem[\protect\citeauthoryear{Meng}{Meng}{1994}]{1994:meng}
Meng, X. (1994).
\newblock Multiple-imputation inference with uncongenial sources of input.
\newblock {\em Statistical Science\/}~{\em 9}, 538 -- 73.

\bibitem[\protect\citeauthoryear{{National Research Council}}{{National
  Research Council}}{2010}]{NRC:2010}
{National Research Council} (2010).
\newblock {\em The prevention and treatment of missing data in clinical
  trials}.
\newblock The National Academies Press: Washington, DC.

\bibitem[\protect\citeauthoryear{Phillips, {Abellan-Andres}, Soren, Bretz,
  Fletcher, France, Garrett, Harris, Kjaer, Keene, Morgan, and andJames
  Roger}{Phillips et~al.}{2016}]{phillips:2016}
Phillips, A., J.~{Abellan-Andres}, A.~Soren, F.~Bretz, C.~Fletcher, L.~France,
  A.~Garrett, R.~Harris, M.~Kjaer, O.~Keene, D.~Morgan, and M.~O. andJames
  Roger (2016).
\newblock Estimands: discussion points from the psi estimands and sensitivity
  expert group.
\newblock {\em Pharmaceutical Statistics\/}.

\bibitem[\protect\citeauthoryear{Ratitch and {O'Kelly}}{Ratitch and
  {O'Kelly}}{2011}]{ratitch:2011}
Ratitch, B. and M.~{O'Kelly} (2011).
\newblock Implementation of pattern-mixture models using standard {SAS/STAT}
  procedures.
\newblock In {\em in Proceedings of PharmaSUG 2011 (Pharmaceutical Industry SAS
  Users Group)}, SP04, Nashville.

\bibitem[\protect\citeauthoryear{Ratitch, O'Kelly, and Tosiello}{Ratitch
  et~al.}{2013}]{ratitch:2013}
Ratitch, B., M.~O'Kelly, and R.~Tosiello (2013).
\newblock Missing data in clinical trials: from clinical assumptions to
  statistical analysis using pattern mixture models.
\newblock {\em Pharmaceutical Statistics\/}~{\em 12}, 337 -- 47.

\bibitem[\protect\citeauthoryear{Rubin}{Rubin}{1987}]{rubin:1987}
Rubin, D. (1987).
\newblock {\em Multiple Imputation for Nonresponse in Surveys}.
\newblock Wiley: New York.

\bibitem[\protect\citeauthoryear{Schafer}{Schafer}{1997}]{schafer:1997}
Schafer, J.~L. (1997).
\newblock {\em Analysis of Incomplete Multivariate Data}.
\newblock Chapman Hall, London.

\bibitem[\protect\citeauthoryear{Siddiqui, Hung, and O'Neill}{Siddiqui
  et~al.}{2009}]{aiddiqui:2009}
Siddiqui, O., J.~H.~M. Hung, and R.~O'Neill (2009).
\newblock {MMRM} vs. {LOCF}: A comprehensive comparison based on simulation
  study and 25 {NDA} datasets.
\newblock {\em Journal of Biopharmaceutical Statistics\/}~{\em 19}, 227--46.

\bibitem[\protect\citeauthoryear{Tang}{Tang}{2015}]{tang:2015}
Tang, Y. (2015).
\newblock An efficient monotone data augmentation algorithm for {B}ayesian
  analysis of incomplete longitudinal data.
\newblock {\em Statistics and Probability Letters\/}~{\em 104}, 146 -- 52.

\bibitem[\protect\citeauthoryear{Tang}{Tang}{2016}]{2016:tang}
Tang, Y. (2016).
\newblock An efficient monotone data augmentation algorithm for multiple
  imputation in a class of pattern mixture models.
\newblock {\em Journal of Biopharmaceutical Statistics\/}.

\bibitem[\protect\citeauthoryear{{van Buuren}}{{van
  Buuren}}{2007}]{Buuren:2007}
{van Buuren}, S. (2007).
\newblock Multiple imputation of discrete and continuous data by fully
  conditional specification.
\newblock {\em Statistical Methods in Medical Research\/}~{\em 16}, 219 -- 42.

\end{thebibliography}
 
\end{document}